\def \be {\begin{equation}}
\def \ee {\end{equation}}
\def \bea {\begin{eqnarray}}
\def \eea {\end{eqnarray}}
\def \nn {\nonumber}

\def \del {\partial}
\def \dels {\partial\kern-.5em / \kern.5em}
\def \As {{A\kern-.5em / \kern.5em}}
\def \Ds {D\kern-.7em / \kern.5em}

\def \dag {\dagger}

\def \d {\delta}

\def \lam {\lambda}

\def \s {\sigma}

\def \th {\theta}

\def \tht {\tilde{\theta}}
\def \Bt {\tilde{B}}
\def \Fh {\hat{\cal F}}

\documentclass[12pt]{article}

\begin{document}
\begin{titlepage}

\begin{center}
\hfill hep-th/0103024\\
\vskip .5in

\textbf{\large Making Non-Associative Algebra Associative
}

\vskip .5in
{\large Pei-Ming Ho}
\vskip 15pt

{\small \em Department of Physics, National Taiwan
University, Taipei 106, Taiwan, R.O.C.}

\vskip .2in
\sffamily{
pmho@phys.ntu.edu.tw}

\vspace{60pt}
\end{center}
\begin{abstract}

Based on results about open string correlation functions,
a nonassociative algebra was proposed in a recent paper
for D-branes in a background with nonvanishing $H$.
We show that our associative algebra obtained
by quantizing the endpoints of an open string
in an earlier work can also be used to reproduce
the same correlation functions.
The novelty of this algebra is that
functions on the D-brane do not form a closed algebra.
This poses a problem to define gauge transformations
on such noncommutative spaces.
We propose a resolution by generalizing
the description of gauge transformations
which naturally involves global symmetries.
This can be understood in the context of matrix theory.

\end{abstract}
\end{titlepage}
\setcounter{footnote}{0}

\section{Introduction} 

In an interesting paper of Cornalba and Schiappa \cite{CS},
they calculated the n-point functions for
open strings ending on a D-brane in
a NS-NS $B$ field background with $H\neq 0$.
From the correlation functions they tried to extract information
about the algebra of functions on the D-brane worldvolume.
They found that Kontsevich's formal expression for
the $\ast$-product can be used to reproduce
the correlation functions.
However, this product is nonassociative
because $B$ is not symplectic when $H\neq 0$.

On the other hand, in an earlier paper \cite{HY},
we derived an associative algebra for
the D-brane worldvolume by quantizing
open strings ending on a D-brane
in curved space with a nontrivial B field background.
The question is whether our algebra can also reproduce
the correlation functions.

When one tries to extract the algebra of functions
on the D-brane from the correlation functions,
the answer is not unique,
because the correlation functions are just numbers,
which we want to interpret as the integrals of functions
on a noncommutative space.
Since we only have information about the integrals of functions,
instead of the functions themselves,
we can not directly obtain the algebra without ambiguity.
It is possible to have many different algebras
that reproduce the same correlation functions after integration.

On the other hand, if we try to derive the algebra
of functions by quantizing an open string on the D-brane,
the result is always an associative algebra.
We should just interpret the algebra of
the endpoint coordinates $X$ and momentum $P$
as the algebra of functions and derivatives on the D-brane.

It is well known that for a constant $B$ field background,
quantization of open string coordinates \cite{CHNC}
and calculation of correlation functions \cite{Schom}
give the same noncommutativity of D-brane worldvolume.
It would be nice to have this kind of agreement for
a generic background.

In this paper, we show that the algebra of \cite{HY}
can also reproduce the open string n-point functions,
but it has the merit of being associative.
The novel property of this algebra is that
the algebra of functions and the algebra of derivatives are mixed up.
The functions do not form a closed algebra by themselves.
This makes it hard to formulate a gauge theory
on such noncommutative spaces.
In the end of this paper we propose a way
to generalize the notion of gauge transformations
for such noncommutative spaces.
It naturally includes a description of global symmetries,
and is reminiscent of the situation in
matrix compactifications.

\section{The Non-Associative Algebra}

It is well known that the noncommutative algebra
\be
[ x^a, x^b ]=i\th^{ab}
\ee
for a constant anti-symmetric tensor $\th$
can be realized on classical commutative functions
by the $\ast$-product
\be \label{star}
f\ast g = fg+\frac{i}{2}\th^{ab}(\del_a f)(\del_b g)
-\frac{1}{8}\th^{ac}\th^{bd}(\del_a\del_b f)(\del_c\del_d g) 
+{\cal O}(\th^3).
\ee

For a generic Poisson structure $\tht(x)$,
the Kontsevich formula \cite{Kon} gives
\bea \label{bullet}
f\bullet g &=& fg+\frac{i}{2}\tht^{ab}(\del_a f)(\del_b g)
-\frac{1}{8}\tht^{ac}\tht^{bd}(\del_a\del_b f)(\del_c\del_d g) \nn \\
& & -\frac{1}{12}\tht^{ad}(\del_d\tht^{bc})
\left( (\del_a\del_b f)(\del_c g)-(\del_b f)(\del_a\del_c g)\right)
+{\cal O}(\tht^3).
\eea

When the field strength
\be
H_{abc}=(\del_a\Bt_{bc})+(\del_b\Bt_{ca})+(\del_c\Bt_{ab})
\ee
for the NS-NS $B$ field background $\Bt$ vanishes,
$\Bt$ defines a symplectic structure on the D-brane
and its inverse gives the Poisson structure
\be \label{thB}
\tht=\Bt^{-1}
\ee
which defines via (\ref{bullet})
the noncommutativity of the D-brane worldvolume
in the zero slope limit of Seiberg and Witten \cite{SW}.

For a more general matrix of functions $\tht$
which is not a Poisson structure,
correpsonding to the case $H\neq 0$,
the algebra defined by the Kontsevich formula is not associative.
The nonassociativity is
\be \label{nonass}
(f\bullet g)\bullet h - f\bullet (g\bullet h) =
\frac{1}{6}K^{abc}(\del_a f)(\del_b g)(\del_c h)+\cdots,
\ee
where
\be \label{K}
K^{abc}=\tht^{ad}(\del_d\tht^{bc})
=\tht^{ad}\tht^{be}\tht^{cf}H_{def}.
\ee
Throughout this paper we will only keep terms
up to the first order of $K$ and to the 2nd order of $\tht$.

For simplicity, let us consider the case
\be \label{Bt}
\Bt_{ab} = B_{ab}+\frac{1}{3}H_{abc}x^c+\cdots,
\ee
where $B$ and $H$ are constant anti-symmetric tensors.
This is the same case considered in \cite{CS}.
We refer to \cite{CS} for discussions on
the effect of nonzero $H$ on the curvature,
as well as many of our conventions and notations.

The last term in the Kontsevich formula (\ref{bullet})
vanishes for the choice (\ref{Bt})
because its coefficient is proportional to
the anti-symmetric tensor $K^{abc}$,
whose indices are contracted with derivatives on $f$ or $g$.
The formula (\ref{bullet}) is simplified to
\bea
f\bullet g &=& fg+\frac{i}{2}\tht^{ab}(\del_a f)(\del_b g)
-\frac{1}{8}\tht^{ac}\tht^{bd}(\del_a\del_b f)(\del_c\del_d g)+\cdots\nn\\
&=& f\ast g -\frac{i}{6}K^{abc}y_c\ast(\del_a f)\ast(\del_b g)+\cdots,
\label{bs}
\eea
where
\be
y_a=B_{ab}x^b,
\ee
and the $\ast$-product is defined by (\ref{star})
for $\th=B^{-1}$.
In deriving (\ref{bs}) we used the relation
\be
\tht^{ab}=\th^{ab}-\frac{1}{3}K^{abc}y_c+\cdots
\ee
which follows from (\ref{thB}) and (\ref{Bt}).
Note that the last term in (\ref{bs})
can also be written as
\be
-\frac{i}{6}K^{abc}(\del_a f)\ast(\del_b g)\ast y_c
\ee
because
\bea
y_a\ast f &=& y_a f + \frac{i}{2}(\del_a f), \\
f\ast y_a &=& y_a f - \frac{i}{2}(\del_a f), \\
\eea
and $K$ is totally antisymmetric.

From (\ref{bs}) it is straightforward to calculate
\be
(\cdots((f_1\bullet f_2)\bullet f_3)\cdots\bullet f_n) =
f_1\ast f_2\ast\cdots\ast f_n+\sum_{i<j}V_{ij},
\ee
where
\be
V_{ij}=-\frac{i}{6}K^{abc}y_c\ast
f_1\ast\cdots\ast(\del_a f_i)\ast\cdots
\ast(\del_b f_j)\ast\cdots\ast f_n.
\ee
Similarly,
\be
f_1\bullet(f_2\bullet\cdots(f_{n-1}\bullet f_n)\cdots)
=f_1\ast f_2\ast\cdots\ast f_n+\sum_{i<j}V'_{ij},
\ee
where
\be
V'_{ij}=-\frac{i}{6}K^{abc}
f_1\ast\cdots\ast(\del_a f_i)\ast\cdots
\ast(\del_b f_j)\ast\cdots\ast f_n\ast y_c.
\ee

It was shown in \cite{CS} that
for open strings ending on a D-brane
in this $B$ field background,
the 2-point function is given by
\be
\int f\bullet g+
\int \frac{1}{3}B_{bc}K^{abc}y_a\ast f\ast g+\cdots.
\ee
The last term is the contribution from
contracting two fields in the interaction term
of the open string action,
and it has the same form
\be \label{V}
\int \frac{1}{3}B_{bc}K^{abc}y_a\ast f_1\ast f_2\ast\cdots f_n
\ee
for $n$-point functions.
They can be taken care of by a modification of the measure
for all correlation functions \cite{CS}
\be
\int F \rightarrow \int (1+\frac{1}{3}B_{bc}K^{abc}y_a)\ast F,
\ee
so we will ignore such terms from now on.

The 3-point correlation function
(up to a term of the form (\ref{V}))
is reproduced by the integral
\be
\int (f\bullet g)\bullet h = \int f\bullet(g\bullet h).
\ee
The nonassociativity of the $\bullet$-product
does not affect the 3-point function.

More generally, we have
\bea
\int (\cdots(f_1\bullet f_2)\cdots\bullet f_n)
&=& \int (f_1\bullet\cdots(f_{n-1}\bullet f_n)\cdots) \nn\\
&=& \int (f_1\ast\cdots\ast f_n + \sum_{i<j}V_{ij}).
\eea
However, for $n$-point functions with $n>3$
one has to use a linear combination of
the $\bullet$-products with various orderings,
weighed by coefficients depending on the modules.

\section{The Associative Algebra}

In \cite{HY}, the symplectic structure for
the endpoint coordinates
of an open string ending on a D-brane
in a background with nonvanishing $H$
is derived in the low energy limit.
Inverting the sympletic two-form for $X$ and $(\del_\s X)$,
we find the Poisson brackets for the coordinates $x$ at $\s=0$
and the momentum density at $\sigma=0$ times $\pi$
(which is the same as the total momentum in our approximation)
\be
p_a=\pi\tht^{-1}_{ab}{X'}^b+\cdots
\ee
as
\bea
(x^a, x^b)&=& \tht^{ab}-\frac{1}{3}K^{abc}p_c+\cdots, \label{xx}\\
(x^a, p_b)&=& \d^a_b+\frac{1}{6}\tht^{-1}_{bc}K^{acd}p_d+\cdots, \\
(p_a, p_b)&=& 0+\cdots, \label{pp}
\eea
where we denote $\Fh^{-1}$ by $\tht$
and keep only terms up to the first order in $K$.
($K$ is defined in (\ref{K}).)
One can check that all Jacobi identities are satisfied
up to our approximation.

A peculiar property of this algebra is that
the commutator of two functions of $x$ is not
a function of $x$ only, but a function of both $x$ and $p$.
Another interesting charater of this algebra is that
it is impossible to realize $p$ by a function $f$ of $x$
such that $(p,x)=(f,x)$.
That is, the derivative $p$ is not an ``inner derivative''.
These properties are the algebraic manifestations
of the fact that $H\neq 0$.

Poisson brackets turn into commutation relations
upon quantization
\be
[f, g]=i(f, g)+\cdots.
\ee
The idea about $\ast$-product is that
the quantum algebra can be realized
on commutative functions by defining a new product
\be
f\diamond g = fg+\frac{i}{2}(f,g)+\cdots.
\ee

If $F=0$, $\Fh$ is just the $B$ field background
in the Seiberg-Witten limit \cite{SW}.
Consider the case $\Fh=\Bt$ given by (\ref{Bt})
in the previous section.
We find that (\ref{xx}) differs from what we get
from (\ref{bullet}) by the term linear in $p$.
Thus we should define a new product
\be
f\diamond g = f\bullet g
-\frac{i}{6}K^{abc}(\del_a f)(\del_b g)p_c+\cdots
\ee
to account for the last term in (\ref{xx}).
This new term that modifies the $\bullet$-product
is precisely what is needed to make it associative
\be
(f\diamond g)\diamond h = f\diamond (g\diamond h).
\ee
Note that here we choose to use $p$ as a c-number,
and $p$ is a derivative in the sense that
$x^a\diamond p_b-p_b\diamond x^a\simeq i\delta^a_b+\cdots$.

Now we have to check that this new product
generates the same 2-point function after integration.
But how do we integrate the terms involving derivatives?
Consider the space of all functions of $x$ as the Hilbert space
of a one-particle system.
Any state in the Hilbert space can be obtained by
acting a function of $x$ on the ``vacuum'' $\rangle$,
which corresponds to the constant function.
The action of $p$ on a state
is determined by the commutation relations between $p$ and $x$,
in addition to the action of $p$ on $\rangle$ give by
\be \label{del0}
p_a\rangle=0.
\ee
Now we define the integration of $f(x,p)$
by $\langle f(x,p)\rangle$,
where $\langle$ is the Hermitian conjugate of $\rangle$.
In our notation, the inner product
$\langle f|g\rangle=\langle f^{\dag}g\rangle$,
and so $\langle f(x)\rangle=\langle 1|f(x)\rangle$,
which is just the integration of $f(x)$ when $\tht=0$.

The translational invariance of the integration
is guaranteed by (\ref{del0}).
To calculate the integral,
one simply commutes $p$ to the right to annihilate $\rangle$,
so that it turns into an integration of a pure function of $x$.
This kind of definition for integrals on noncommutative spaces
has been used in many occasions \cite{integral}.

One can check that
\be
\langle f\diamond g\rangle = \int f\bullet g,
\ee
and so the new product is consistent with the 2-point function.

For the 3-point function, we find
\be
f\diamond g\diamond h = f\bullet(g\bullet h)-\frac{i}{6}K^{abc}
[(\del_a f)(\del_b g)h+(\del_a f)g(\del_b h)+f(\del_a g)(\del_b h)]p_c
+\cdots,
\ee
and so
\be
\langle f\diamond g\diamond h\rangle = \int f\bullet(g\bullet h).
\ee
Thus we check that the new $\diamond$-product also
reproduces the 3-point function correctly.
What we gained by the modification to the $\bullet$-product
is that our algebra is now associative.

Denote the insertion coordinates
at the boundary of the open string by $\tau_i$.
The 4-point function depends on the module $m$
which is the cross ratio
\be
m=(\tau_4-\tau_3)(\tau_2-\tau_1)
(\tau_4-\tau_2)^{-1}(\tau_3-\tau_1)^{-1}.
\ee
The integral
\be
\langle f_1\diamond\cdots\diamond f_4\rangle =
\int\left( f_1\ast\cdots\ast f_4+\sum_{i<j}V_{ij} \right)
\ee
properly reproduces the 4-point function \cite{CS}
except the term depending on the module
\be
\frac{1}{6}L(m)K^{abc}\int
f_1(\del_a f_2)(\del_b f_3)(\del_c f_4),
\ee
where $L(m)$ is a function ranging between $0$ and $1$ \cite{CS}.
One should integrate over $m$
to make connection with the D-brane action.
This term can be taken care of separately by a term of the form
\be
K^{abc}\int
f_1\diamond(\del_a f_2)\diamond(\del_b f_3)\diamond(\del_c f_4).
\ee

In general,
\be
\langle f_1\diamond\cdots\diamond f_n\rangle =
\int (f_1\ast\cdots\ast f_n+\sum_{i<j}V_{ij}).
\ee
To take care of the module dependent terms of the $n$-point functions,
one has to superpose $\bullet$-products of various orderings \cite{CS}.
For the $\diamond$-product,
one has to take care of these terms separately.
Since we do not have to use more $\diamond$-products
to write down an $n$-point function,
the $\diamond$-product is as good as the $\bullet$-product
for the purpose of generating correlation functions.
It is also possible that we do not have to
worry about the 4-point functions because
the quartic terms in the low energy D-brane action
may be uniquely determined by gauge symmetry
after the quadratic and cubic terms are given.
The benefit of using the $\diamond$-product is of course
that it is associative, and so one can use it to
formulate algebraic structures of the theory such as symmetries.

\section{Noncommutative Gauge Theory}

In this section we discuss how to construct a gauge theory on
the noncommutative space defined by (\ref{xx})-(\ref{pp}).
This noncommutative space is quite different from
the cases with $H=0$ in that
the functions of $x$ do not form a closed subalgebra.
When one makes a gauge transformation on
a field $\phi(x)$ in the adjoint representation
\be \label{gauge}
\phi\rightarrow \phi'=U\diamond\phi\diamond U^{\dagger},
\ee
the result $\phi'$ is generically not a function of $x$
although $\phi$ and $U$ are functions of $x$ only.
This poses a serious problem in
constructing a gauge field theory.

We propose that the natural notion of gauge transformations here
is to restrict $U$ to be functions of $x$ and $p$ such that
$\phi'$ will be a function of $x$ for any $\phi$.
From the viewpoint of the matrix model,
both $x$ and $p$ have the same origin
as (infinite dimensional) matrices.
Note that in the classical case this notion of gauge symmetry
does not completely coincide with the usual definition.
We shall further restrict the symmetry so that
it has the correct classical limit when $\tht\rightarrow 0$.
In fact, the notion of gauge symmetry has already been
generalized for compactified matrix theory \cite{HW1,HW2,HW3}.
It has been pointed out that both global symmetries
and gauge symmetries for the theory after compactification
come from gauge symmetries of the original matrix theory \cite{HW2,HW3}.
For example, the quotient condition for the matrix model
compactified on a torus \cite{GRT} is
\be
U^{\dag}_j X_i U_j = X_i + 2\pi\delta_{ij}R_j.
\ee
For the dual theory in which
$X_i$ is identified with the covariant derivative,
\be
X_i\rightarrow u^{\dag}X_i u
\ee
is a gauge transformation if $u U_i = U_i u$.
It is a global (translational) symmetry
if $u U_i = q_i U_i u$ for some phase factor $q_i$ \cite{HW2}.
The symmetries of the compactified theory
are given by any unitary transformation
that commutes with the quotient conditions.
Our proposal is in the same spirit.

For a field $\phi(x)$ in the adjoint representation,
an infinitesimal gauge transformation by $\lam$ is
\be
\delta\phi(x) = [\lam, \phi(x)].
\ee
For $\lam$ to make a valid transformation,
we require that $\delta\phi$ is a function of $x$
for any function $\phi(x)$.
Expanding $\lam(x,p)$ in powers of $p$
\be
\lam = \lam_0(x)+\lam_1^a(x)p_a+\lam_2^{ab}(x)p_a p_b+\cdots,
\ee
where $\lam_2^{ab}=\lam_2^{ba}$,
we find that the requirement is matched if
\bea
\lam_1^a   &=&-2\th^{ab}(\del_b\lam_0)+\th^{ab}\zeta_b, \\
\lam_2^{ab}&=&\th^{ac}\th^{bd}((\del_c\del_d\lam_0)+\xi_{cd}),
\eea
where $\zeta$ satisfies
\be \label{zeta}
(\del_a\zeta_b)-(\del_b\zeta_a) = \frac{1}{3}H_{abc}\th^{cd}\zeta_d,
\ee
and $\xi$ is defined by
\be
\xi_{ab}=-\frac{1}{4}((\del_a\zeta_b)+(\del_b\zeta_a)).
\ee
It follows from this that
\be \label{transf}
\delta\phi = -i\th^{ab}(\del_a\lam_0)(\del_b\phi)
+i\th^{ab}\zeta_a(\del_b\phi) + \cdots
\ee
is a function of $x$ only.

It is interesting to note that
$\zeta$ defines the Killing vector for $\tht$.
Eq. (\ref{zeta}) is precisely
the constraint for the coordinate transformations
\be
\delta x^a=\th^{ab}\zeta_b
\ee
which keep $\tht$ invariant at $x=0$,
and the contribution of $\zeta$ to $\delta\phi$
in (\ref{transf}) is exactly this coordinate transformation.
This means that the degrees of freedom in $\zeta$
correspond to global symmetries,
and those in $\lam(x)$ correspond to gauge transformations.
As we mentioned earlier,
this is in close analogy with
the situation in matrix compactifications.

For the gauge potential $A$, we can define its
gauge transformation via the covariant derivative.
If we define the covariant derivative by $D=p+A(x)$,
we will find that it is not of the same form
after a gauge transformation.
One simple way out is to define it by $D=x+A$,
which appears naturally in the context of matrix model
\cite{Li,SW,Corn}.
Requiring $D$ to transform in the adjoint representation
determines the gauge transformation of $A$
\be
\delta A^a = [\lam, D^a].
\ee

For a field $\psi(x)$ in the fundamental representation,
we can define its gauge transformation ($\zeta=0$)
for $\psi\rangle$ as
\bea
\delta\psi\rangle &=& \lam\diamond\psi\rangle \nn\\
&=& \left(\lam_0\psi-
\frac{i}{2}\th^{ab}(\del_a\lam_0)(\del_b\psi)\right)\rangle
+\cdots.
\eea
If we consider $\lam\diamond\psi$ without acting on $\rangle$,
there will also be terms depending on $p$.
Thus when trying to write down a gauge invariant action,
we can have terms like $\langle\psi^{\dag}\psi\rangle$,
but not terms like
$\langle(\psi^{\dag}\psi)^2\rangle$.

\section*{Acknowledgment}

The author thanks Miao Li and Hyun Seok Yang
for helpful discussions.
This work is supported in part by
the National Science Council, Taiwan, R.O.C.
the Center for Theoretical Physics
at National Taiwan University,
and the CosPA project of the Ministry of Education, Taiwan.

\end{document}